\documentclass[conference]{IEEEtran}
\IEEEoverridecommandlockouts
\usepackage{cite}
\usepackage[caption=false,font=normalsize,labelfont=sf,textfont=sf]{subfig}
\usepackage{amsmath,amssymb,amsfonts}
\usepackage{algorithmic}
\usepackage{graphicx}
\usepackage{textcomp}
\usepackage{xcolor}
\usepackage{stfloats}
\usepackage[numbers,sort&compress]{natbib}

\def\BibTeX{{\rm B\kern-.05em{\sc i\kern-.025em b}\kern-.08em
    T\kern-.1667em\lower.7ex\hbox{E}\kern-.125emX}}
\begin{document}

\title{Map-assisted Wireless TDOA Localization Enhancement Based On CNN}

\author{

\IEEEauthorblockN{Yiwen Chen\textsuperscript{1}, Tianqi Xiang\textsuperscript{1}, Xi Chen\textsuperscript{2}, Xin Zhang$^{1\ast}$ \thanks{\textcolor{red}{This work has been accepted for presentation at the 2024 IEEE 6th Advanced Information Management, Communicates, Electronic and Automation Control Conference(IMCEC 2024).}}
}
\IEEEauthorblockA{
	1. School of Information and Communication Engineering, Beijing University of Posts and Telecommunications, Beijing, China\\
	2. National Research Center for Information Science and Technologies, Tsinghua University, Beijing, China\\
	chenyiwen@bupt.edu.cn, xiangtianqi@sina.com, chenxiee@tsinghua.edu.cn, zhangxin@bupt.edu.cn\\
	Corresponding Author: Xin Zhang \  Email: zhangxin@bupt.edu.cn. 
}






}

\maketitle
\begin{abstract}
For signal processing related to localization technologies, non line of sight (NLOS) multipaths have a significant impact on the localization error level. This study proposes a localization correction method based on convolution neural network (CNN), which extracts obstacle features from maps to predict the localization errors caused by NLOS effects. A novel compensation scheme is developed and structured around the localization error in terms of distance and azimuth angle predicted by the CNN. Four prediction tasks are executed over different building distributions within the maps for typical urban scenario, resulting in CNN models with high prediction accuracy. Finally, a thorough comparison of the accuracy performance between the time difference of arrival (TDOA) localization algorithm and the results after the error compensation reveals that, generally, the CNN prediction approach demonstrates great localization error correction performance, improving TDOA accuracy by 75\%. It can be observed that the powerful feature extraction capability of CNN can be exploited by processing surrounding maps to predict the localization error distribution, showing great potential for further enhancement of TDOA performance under challenging scenarios with rich multipath propagation.
\end{abstract}


\section{Introduction}
The localization of user equipment (UE) is a pivotal feature in the era of 6G. Time difference of arrival (TDOA) is a simple and efficient wireless localization method that relies on the time it takes for signals from a single terminal to reach multiple monitoring stations or base stations (BSs). TDOA utilizes time-distance conversion for intersection localization, a method widely adopted due to its modest computational requirements \cite{b1}. In scenarios with a direct Line of Sight (LOS) between the transmitter (Tx) and receiver (Rx), where the wireless signal travels along a straight path, TDOA exhibits excellent localization performance. However, in Non-Line-of-Sight (NLOS) scenarios, the wireless signal undergoes reflection, diffraction, and scattering, resulting in  localization errors that may exceed 100 meters in specific cases \cite{b2}. 

\setlength{\parindent}{1em}To address multipath-related errors of TDOA in NLOS conditions, various schemes have been introduced to optimize localization accuracy. In \cite{b2}, a three-stage TDOA measurement processing algorithm is proposed to identify and mitigate localization errors. Additionally, more hardware has been utilized to enhance the accuracy of TDOA time synchronization \cite{b3} \cite{b4}. In \cite{b4}, an adaptive robust particle filtering (ARPF) algorithm is proposed to improve the localization accuracy and robustness. \cite{b5} introduces an automatic control extended Kalman filter (ACEKF) localization algorithm optimized for TDOA. Despite these advancements, it is noteworthy that none of the literature above mentions a method that directly extracts or utilizes map information through deep learning to estimate NLOS multipath-related positioning errors.

\setlength{\parindent}{1em}Deep learning (DL) offers a significant advantage in applications that involve data features processing, such as computer vision and signal processing. In \cite{b6}, a long short-term memory (LSTM) network was proposed to determine target state and predict TDOA, addressing issues related to localization measurement errors or missing data. Similarly, \cite{b7} proposed the utilization of Deep Neural Networks (DNN) to predict corrected arrival times, indirectly reducing TDOA errors. Both \cite{b6} and \cite{b7} optimized arrival time error information based on DL, leading to an indirect improvement in localization accuracy. In \cite{b8}, a direct position estimation method was introduced using Convolutional Neural Networks (CNN) and channel impulse response (CIR) alongside ground truth localization data, analyzing the signal's time series for multipath effects without directly utilizing terrain information for position estimation. \cite{b9} proposed a new localization technique based on DNN capable of learning the complex mapping between TDOA measurements obtained from multiple BSs and the 3D position of UE, considering the connections between the TDOA distribution and the propagation environment. In \cite{b10}, a CNN-based fingerprinting for localization was analyzed, utilizing a realistic ray-tracing-based channel state information (CSI) containing environmental information to enhance localization accuracy. As the environmental information contains characteristics of NLOS distribution \cite{b9} \cite{b10}, there is the potential to extract and utilize this information to improve TDOA accuracy. However, in \cite{b6}-\cite{b10}, maps containing environment information are not directly used for location estimation or prediction, but are indirectly employed through factors such as CIR, CSI and the arrival time, which would have limitations on fully exploiting the information from maps, as the more stages of processing there are, the more loss the localization related information will suffer from. In \cite{b11} and \cite{b12}, it is proposed that the channel features of a certain map can be extracted based on CNN by DL to predict the mmWave CSI. Further, it is expected that DL can be used to predict or refine the localization results of TDOA by extracting multi-path channel information from maps, not just for mmWave. 

\setlength{\parindent}{1em}In this study, we address the challenges of low TDOA accuracy in NLOS conditions, where map data has not been fully utilized. We apply CNN to extract environmental features directly from maps, predicting the TDOA distance-angle-error (DA-error) distribution, as explained in Section II, facilitating error compensation in localization. We utilize localization errors, obtained by comparing TDOA results with ground truth, as training data to establish a CNN model capable of predicting localization error distributions across different maps. The predicted results from this model are then employed to compensate for TDOA errors.

\setlength{\parindent}{1em}The main contributions of this paper are as follows. (1) We present a novel TDOA localization enhancement method with a significant improvement in accuracy. To the best of our knowledge, this map-assisted localization enhancement scheme based on DL has not been reported in the literature. (2) To ensure fidelity of the results, we employ a ray-tracing module to obtain CSI and then derive the time of arrival through link-level simulation, comprehensively modeling building obstructions and reflection effects in the maps. (3) Comprising more than 1500 maps, we create an extensive training dataset to achieve a considerable accuracy improvement and good generalization ability. These maps encompass various NLOS scenarios, including different building quantities, height distributions, and propagation characteristics, covering both single-path and multi-path conditions. In this paper, 'single path' denotes a direct ray between a Rx and Tx, whereas 'multi-path' includes direct and reflected rays between an Rx and Tx. (4) We assess the localization accuracy of the CNN-based method, achieving a root mean square error (RMSE) within 1 m and 0.5 rad across the entire loss distribution. Following DA-error compensation, the localization error is reduced by 75\% compared to the TDOA results without error correction.

\setlength{\parindent}{1em}The rest of this paper is organized as follows. Section II introduces the localization principle of TDOA and its connection with NLOS multipath, revealing the potential for optimization. Section III describes the map-assisted CNN-based localization DA-error compensation scheme. Section IV presents simulation results and discussion. Section V provides the conclusion and future work.

\section{Wireless Localization Scheme}
\setlength{\parindent}{1em}In the localization process under study, only the time difference needs to be measured to obtain the distance difference between UE and BSs \cite{b13}. For TDOA in three-dimensional space, at least four observation BSs are required for localizing a UE \cite{b14}. A typical spatial positional relationship of observation BSs and the target UE is shown in Fig.\ref{fig1}. Assume the spatial position coordinate of each BS in the localization is $\left( {{x_i},{y_i},{z_i}} \right),i = 1,2,3,4$. Number 1 designates the main receiving BS, while the others represent sub-receiving BSs \cite{b14}. The TDOA equation obtained according to the arrival time difference is as follows:

\begin{figure}[t]
\centering
\includegraphics[width=2.3in]{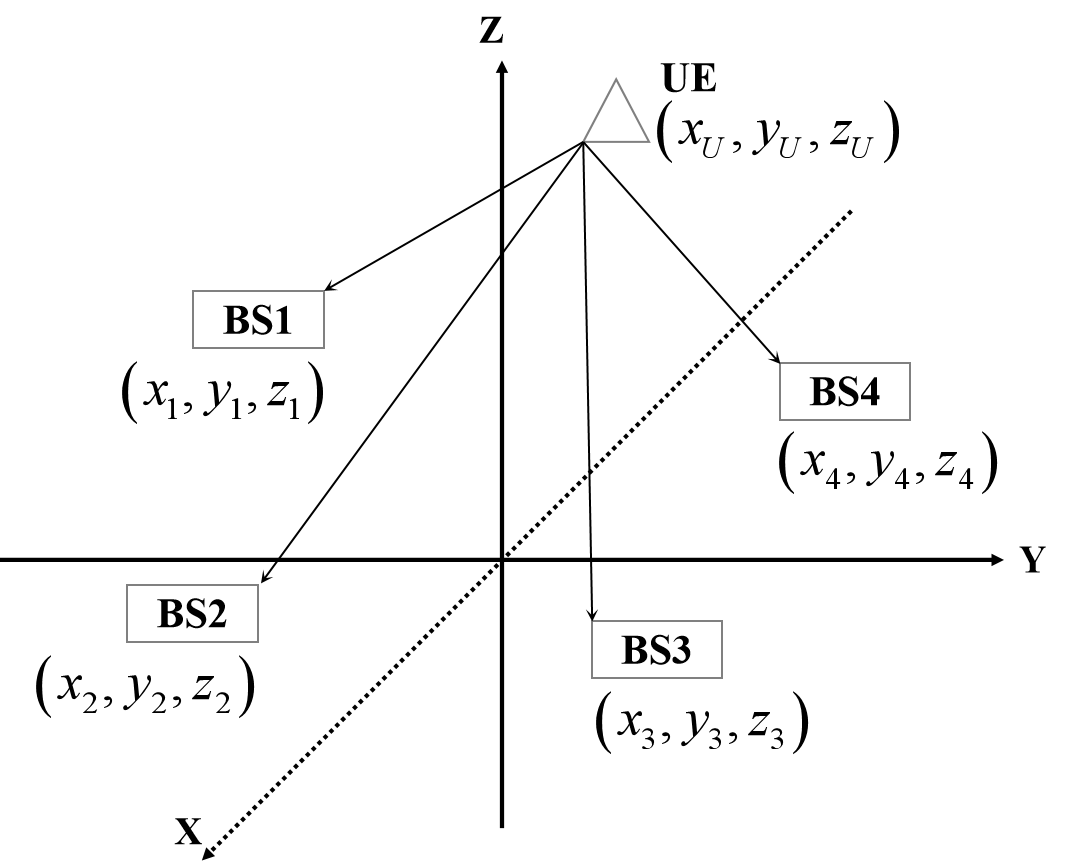}
\caption{Positional relationship of UE and BSs}
\label{fig1}
\end{figure}

\begin{scriptsize} 
\begin{equation}
\begin{aligned}
c{\tau _{1j}} = &\sqrt {{{\left( {{x_1} - {x_U}} \right)}^2} + {{\left( {{y_1} - {y_U}} \right)}^2} + {{\left( {{z_1} - {z_U}} \right)}^2}} - \\
			&\sqrt {{{\left( {{x_j} - {x_U}} \right)}^2} + {{\left( {{y_j} - {y_U}} \right)}^2} + {{\left( {{z_j} - {z_U}} \right)}^2}}
\end{aligned}
\end{equation}
\end{scriptsize} 

\setlength{\parindent}{0cm}where, $c$ is the speed of light, $j=2,3,4$, subscript \textit{U} means UE, $\left( {{x_U},{y_U},{z_U}} \right)$ denotes the location of the target UE and $\tau _{1j}$ is the time difference of arrival between the main receiving BS and sub-receiving BSs. By substituting the time difference of arrival ($\tau$) and the coordinates of 4 BSs into equation(s) (1), the coordinates of the UE can be obtained. Unlike time of arrival (TOA), in TDOA, precise synchronization between BS and UE is not required, as long as the positions of BSs are fixed \cite{b13}. However, due to the influence of NLOS and multi-path effect, the arrival time of the signal can be altered when blocked or reflected by obstacles or scatterers, resulting in a deviation of the arrival time difference. The arrival time difference with NLOS-induced errors can lead to significant localization errors exceeding 100 meters [2], necessitating additional effort to mitigate the impact of NLOS. 

\begin{figure}[t]
\centering
\includegraphics[width=2.2in]{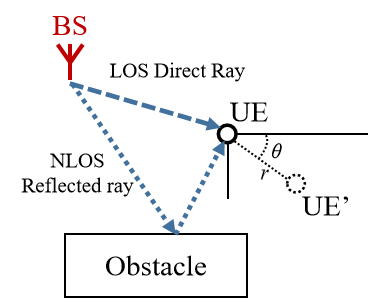}
\caption{A simplified illustration of localization error}
\label{fig2}
\end{figure}

\setlength{\parindent}{1em}As shown in Fig.\ref{fig2}, the localization error is primarily attributed to the multi-path effect in NLOS-dominating environment, where the propagation distance of the reflected ray is longer than that of the direct ray, causing the localization result of the reflected ray to deviate from the true value. As this localization error is closely tied to the distribution of obstacles in the environment, what if the environment-related NLOS pattern could be identified and exploited from maps to directly estimate the localization error? In our previous work \cite{b15}, CNN was proposed to extract environmental features from maps, predicting the distribution of propagation loss related to building distribution. Given that the localization error is also influenced by the distribution of obstacles in the propagation environment, we can process maps to directly estimate the localization error, leveraging our work in \cite{b15} in a novel manner. Fig.\ref{fig2} indicates that the localization deviation is represented by distance \textit{r} and angle $\theta$ in the plane, referred to as DA-error (Distance-Angle-error). In \cite{b16}, it is suggested that distance and angle in polar coordinates can be used for error compensation operations. Intuitively, the location errors for Cardician coordinates would be an array of 2D or 3D vectors over a map, and it would be challenging to discern how they are linked to the spatial distribution of obstacles. Therefore, we adopt distance and azimuth angle to describe the location errors caused by NLOS multipaths, as their spatial distribution features can be extracted by CNN.

Then we introduce a CNN to extract high dimensional characteristics associated with the localization DA-errors in various NLOS environments from corresponding maps. Subsequently, during the TDOA estimation process, the localization DA-error predicted by the CNN is utilized to correct the localization results, thereby improving the accuracy in NLOS scenes.
\section{Localization Correction based on CNN}
\setlength{\parindent}{1em}To apply CNN for compensating localization errors, we propose to commence with predicting the errors associated with direct and reflected rays in NLOS maps. In our previous work \cite{b15}, we have realized the prediction of path loss by processing maps by CNN. In this paper, we leverage CNN to predict the DA-error distribution of TDOA by deeply mining the CNN’s ability to extract map features. Subsequently, the predicted localization DA error is used to correct the TDOA result. To enhance the generalization of the DA-error prediction model, this paper constructs a dataset covering various NLOS scenarios. For achieving online TDOA DA-error correction, we adopt the localization error compensation scheme introduced in \cite{b17} following offline model training.

\subsection{Data Set Construction}

\begin{figure}[t]
\centering
\includegraphics[width=2.8in]{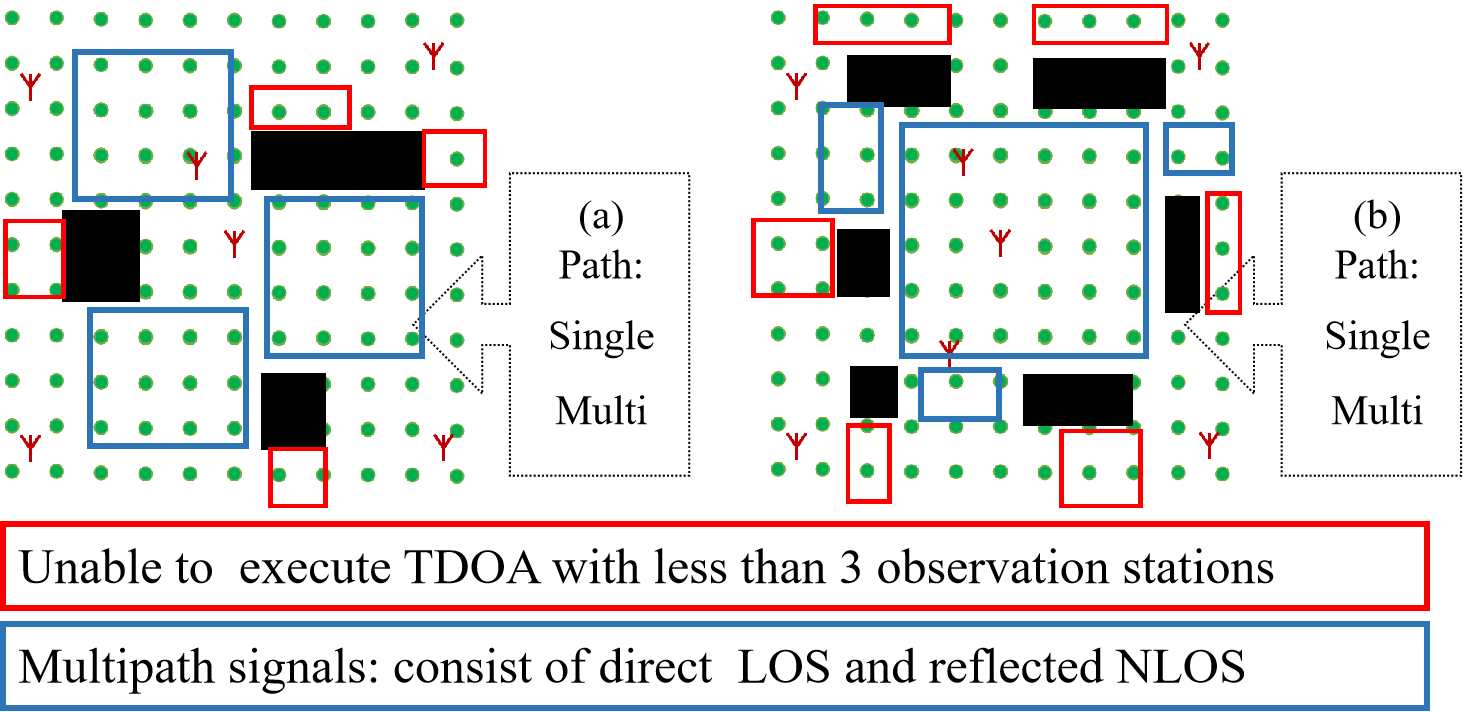}
\caption{Building distributions, (a) 3 buildings, (b) 6 buildings.}
\label{fig3}
\end{figure}

\setlength{\parindent}{1em}To enhance the performance of the predictive model, it is necessary to comprehensively consider the establishment of localization error labels from the perspective of building and observation point distributions, which involves factors determining multi-path effect of NLOS, such as the number, location and height distribution of buildings. As depicted in Fig.\ref{fig3}, the elevation of the building is assumed to be between 5 to 10 m, while that of the open space is 0 m. 3 or 6 buildings are randomly scattered across the map, ensuring the dataset is comprehensive but not overly complicated. Subsequently, the two-dimensional maps of elevation information will be processed as the images input to the CNN. 

\setlength{\parindent}{1em}In Fig.\ref{fig3}, the seven candidate Tx points (green dots) are placed in fixed locations of the map. The 120 Rx points (red dots), where each point includes a distance error and an azimuth angle error, referred to as DA-error in our work, are distributed at equal intervals, serving as the localization DA-error labels for the corresponding map. The heights of Txs are uniformly set to 10 m, while those of Rxs are consistently set to 1 m, which implies that the localization DA-error correction for Rx needs to be conducted only in a 2D plane. Following the principle of TDOA, for each Rx, three reference stations for localization are required among the 7 Tx points to determine the target location on a 2D plane. In the single path case, a direct ray exists  between each pair of Rx and Tx, known as LOS propagation, unless the ray is blocked by the buildings. In the multi-path case, at least one reflected ray introduced by buildings exists among all Rx-Tx pairs, termed NLOS propagation. The optimization focus is on the localization error caused by the reflection propagation of NLOS, with the single-path case serving as a baseline for the multi-path case.

\setlength{\parindent}{1em}As shown in Fig.\ref{fig3}, the area bounded by red boxes represents scenarios where signals from two or fewer observation BSs can reach the localization targets, in either the single-path or multi-path case, rendering the TDOA unfeasible. The localization DA-error label is determined by the difference between the TDOA result and the ground truth. To standardize the scale of the output label, for Rxs where three reference stations cannot be applied for TDOA or Rxs covered by the building, a worst-case value of 50 m is set as the label. Besides, the area bounded by blue boxes models the influence of multi-path propagation in the maps containing both the LOS direct ray and NLOS reflected ray. Considering the random distribution of buildings in terms of both size and location, with the number of buildings set at 3 or 6, along with the propagation paths of single and multiple signals, the dataset is divided into four cases \cite{b18} to train the CNN.

\begin{figure}[t]
\centering
\includegraphics[width=2.4in]{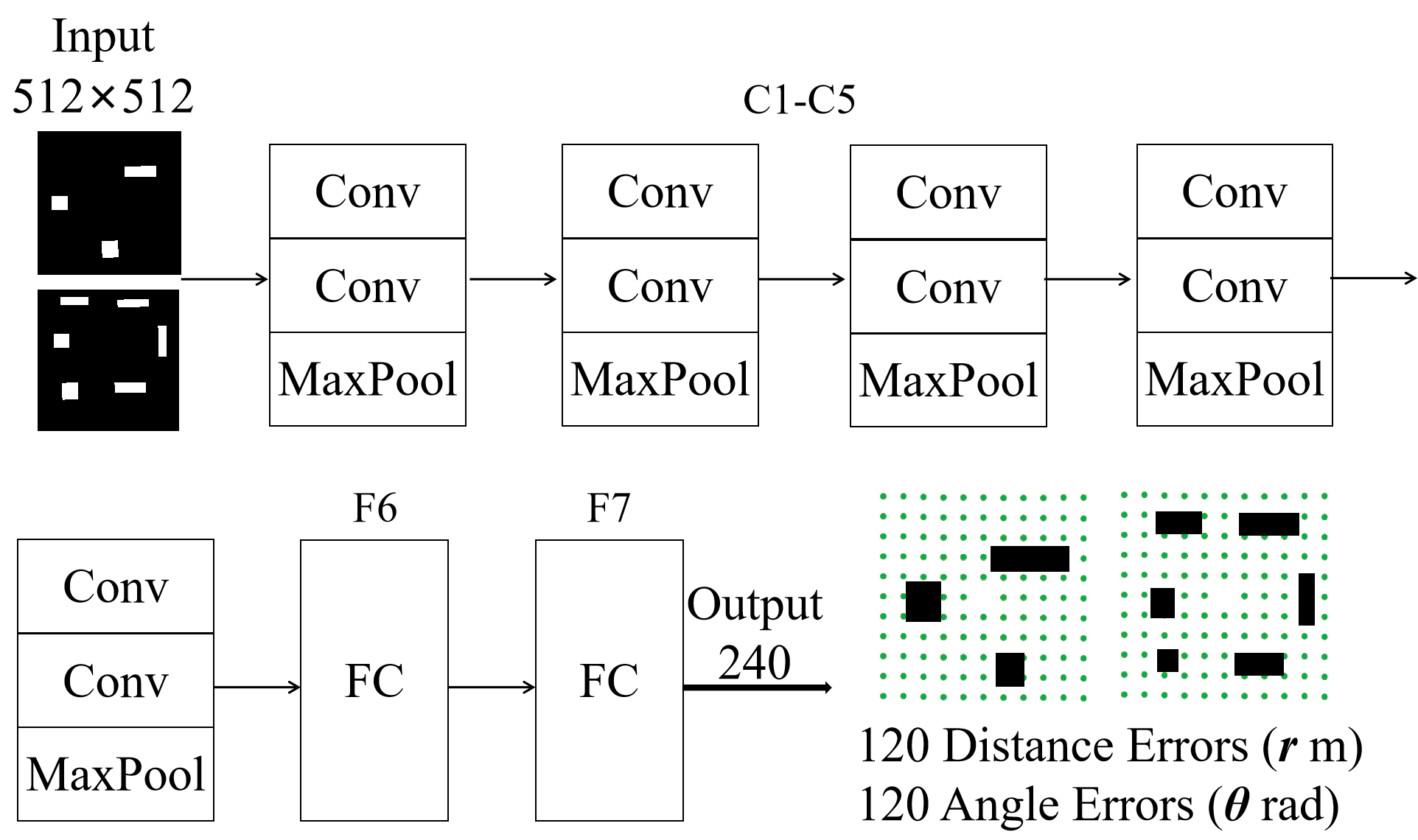}
\caption{Architecture of the CNN to predict localization error}
\label{fig4}
\end{figure}

\begin{figure}[t]
\centering
\includegraphics[width=2.3in]{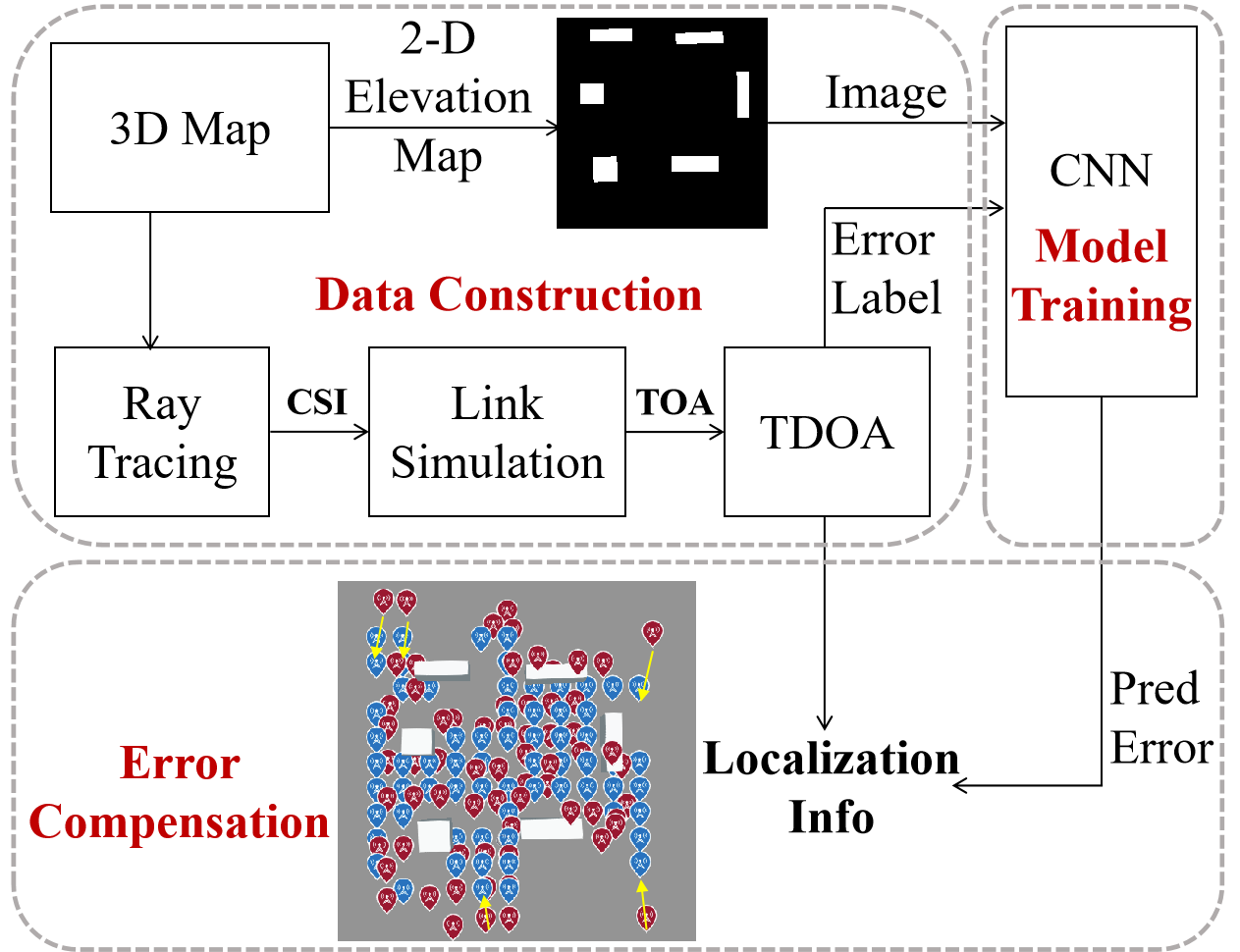}
\caption{Diagram of the localization DA-error correction scheme}
\label{fig5}
\end{figure}

\subsection{Offline Model Training}

\setlength{\parindent}{1em}As CNN should be trained by the localization DA-error of TDOA under certain scenarios, and to maintain generality, randomly generated 3D urban street maps are converted into 2D elevation format as input \cite{b15}. In Fig.\ref{fig4}, the size of the input map is set to 512x512 pixels. For map-based TDOA localization in NLOS, a general CNN architecture, VGG-net \cite{b19} is considered due to its proven modeling capability, where the two dimensional elevation map serves as the model's input and the output is the localization DA-error distribution. The structure we adopt deviates from the typical VGG-net, where the convolution layer uses a 5x5 convolution kernel and has fewer channels. The output of CNN is expected to be 240 predicted localization DA-errors, used for localization DA-error correction, with 120 distance errors and 120 angle errors. With high prediction efficiency, the CNN predicts both distance error and angle error, providing the loss values of 120 distance errors and 120 angle errors in parallel. The loss function is defined as RMSE between the predicted value and the label value, as follows:

\begin{scriptsize} 
\begin{equation}
Los{s_{RMSE}} = \sqrt {\frac{1}{N}\sum\limits_{i = 1}^N {{{\left( {\widetilde {{e_i}} - {e_i}} \right)}^2}} }
\end{equation}
\end{scriptsize} 

\setlength{\parindent}{0cm}where $i$ is the index of a prediction error in a map, ${e_i}$ is the label value, $\widetilde {{e_i}}$ is the the predicted value, and $N$ is 240, including 120 distance errors and 120 angle errors.

\setlength{\parindent}{1em}In Fig.\ref{fig5}, the following steps are used to generate the DA-error labels: (1) CSI is obtained from 3D maps by using a ray tracing algorithm, which includes the delay and complex magnitude response of each propagation path, along with the center carrier frequency of the downlink signal. (2) Time of arrival is calculated from the obtained CSI with a link-level module. This module generates the estimated time of arrival in seconds, by performing a time of arrival simulation using the correlation detection method from the received baseband signal, whose parameters are set according to \cite{b20} as shown in Tab.\ref{Tab1}. (3) According to the time of arrival, the TDOA algorithm is employed to determine the localization result. (4) Based on the difference between the localization result and the ground truth, the localization DA-error distributions in the map are produced as labels.

\begin{table}[t] 
\centering
\caption{Link-level simulation parameters} 
\label{Tab1} 
\begin{tabular}{|c|c|} 
\hline 
\scriptsize{Detecting algorithm}&\scriptsize{first-path detection}\\
\hline 
\scriptsize{Carrier frequency and system bandwidth}&\scriptsize{4 GHz, 400 MHz}\\
\hline 
\scriptsize{Noise factor of the UE}&\scriptsize{9 dB}\\
\hline 
\scriptsize{Range of resource block index}&\scriptsize{1-273}\\
\hline  
\scriptsize{Linear interfere power form other UEs}&\scriptsize{0 W}\\
\hline 
\scriptsize{Relative magnitude threshold}&\scriptsize{-10 dB}\\
\hline
\scriptsize{Num of symbols occupied by reference signal}&\scriptsize{8}\\
\hline
\scriptsize{DL-PRS-CombSizeN}&\scriptsize{2}\\
\hline
\end{tabular} 
\end{table}

\subsection{Online Error Compensation}
\setlength{\parindent}{1em}To validate the proposed scheme, it is initially assumed that 3D map information of the surrounding area is available at the UE. Then, as shown in Fig.\ref{fig5}, the map is converted into a 2D elevation format and fed into the trained CNN model to obtain the localization error distribution over it. Based on the TDOA algorithm, a rough estimate of UE's location is obtained, and the nearest corresponding error value is determined by searching from the predicted location error distribution diagram. Subsequently, the predicted localization error of TDOA is employed for DA-error compensation, covering both distance error ($r$ in meters) and azimuth angle error ($\theta$ in radians), as illustrated in Fig.\ref{fig6}. Since the height of all UEs is the same, only 2D coordinates need to be corrected. Finally, the compensation is carried out as follows:

\begin{figure}[b]
\centering
\includegraphics[width=1.3in]{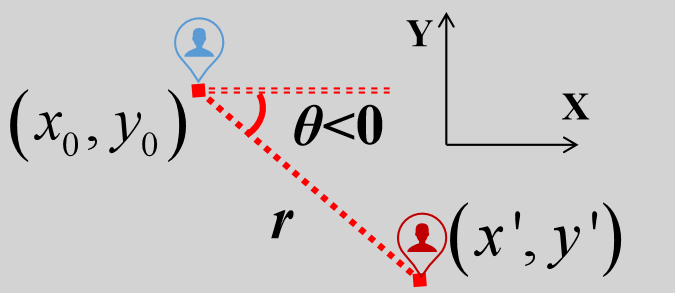}
\caption{DA-error compensation}
\label{fig6}
\end{figure}

\begin{scriptsize} 
\begin{equation}
x = x' - r\cos \left( \theta  \right),y = y' - r\sin \left( \theta  \right),\theta  \in \left[ { - \pi ,\pi } \right]
\end{equation}
\end{scriptsize} 

\setlength{\parindent}{0cm}where $\left({x',y'} \right)$ is the result of TDOA, and $\left({x,y} \right)$ is the location after error compensation. The goal is for the compensated location to be as close as possible to the ground truth value $\left( {x_0, y_0} \right)$ in Fig. \ref{fig6}.

\section{Simulation Results}
\begin{table}[t] 
\begin{center} 
\caption{Comparison of predicted loss in different scenarios} 
\label{Tab2} 
\begin{tabular}{|c|c|c|} 
\hline 
\scriptsize{\textbf{RMSE for Localization}}&\scriptsize{\textbf{Distance (m)}}&\scriptsize{\textbf{Angle (rad)}}\\
\hline 
\scriptsize{\textbf{Single Path in 3BLDG}}&0.4192&0.2981\\
\hline 
\scriptsize{\textbf{Multipath in 3BLDG}}&0.5313&0.3016\\
\hline 
\scriptsize{\textbf{Single Path in 6BLDG}}&0.5705&0.3314\\
\hline 
\scriptsize{\textbf{Multipath in 6BLDG}}&0.6840&0.4086\\
\hline
\end{tabular} 
\end{center} 
\end{table}

\setlength{\parindent}{1em}Based on TDOA in maps with both NLOS and LOS propagation modes, we generate four typical scenarios as outlined in Tab.\ref{Tab2}, to train and evaluate the CNN. Each scenario comprises 1300 training sets, 200 validation sets, and 20 test sets. The CNN hyperparameter settings are as follows: Batch Size set to 4, Initial Learning Rate set to 0.00001, Learning Rate Decay Factor set to 0.95, Epoch set to 200, Learning Rate Decay using Exponential Decay, Optimizer using the Adam method, and Activation Function using the ReLu function. The radio electromagnetic signals are assumed as ideal direct rays or perfectly reflected rays.

\subsection{Predicted Localization Errors based on CNN}
\setlength{\parindent}{1em}Tab.\ref{Tab2} illustrates the loss performance of the trained CNN models for predicting distance errors across different datasets. It is indicated that: (1) The increase in the number of buildings tends to reduce the prediction accuracy when both LOS and NLOS propagation modes are present. (2) Predicting the propagation scenario of multi-path is more challenging compared to that of a single path. (3) The predicted localization DA-error loss is consistently maintained within 1 m and 0.5 rad, reflecting excellent accuracy performance. With an expansion of the training dataset for better generalization, the CNN prediction accuracy could be further enhanced.

\subsection{Results of Localization DA-Error Compensation}
\begin{figure}
\begin{minipage}[b]{1\linewidth}
    \centering
    \subfloat[][\begin{scriptsize}Single Path\end{scriptsize}]{\includegraphics[width=4.2cm]{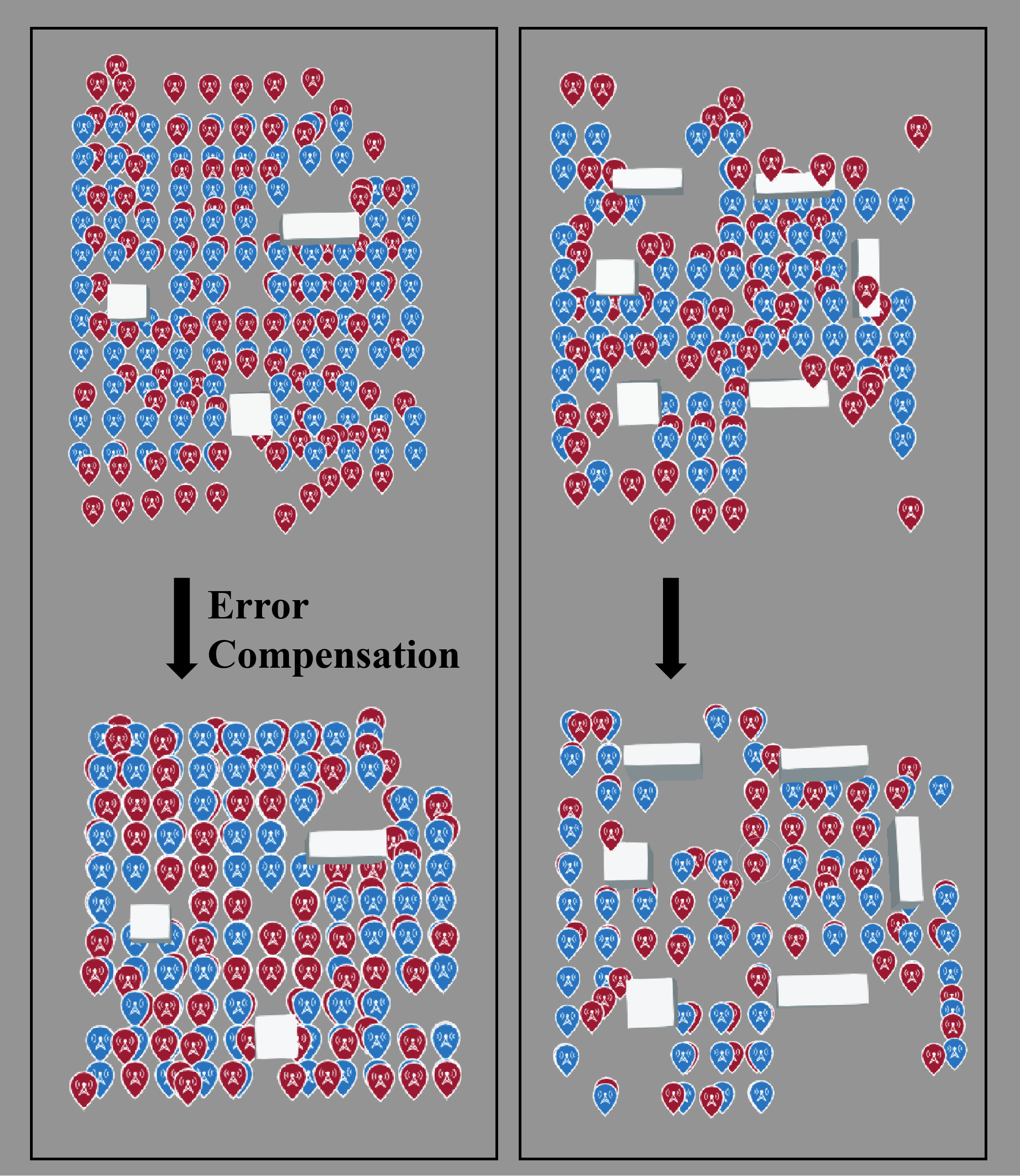}}
    \subfloat[][\begin{scriptsize}Multipath\end{scriptsize}]{\includegraphics[width=4.2cm]{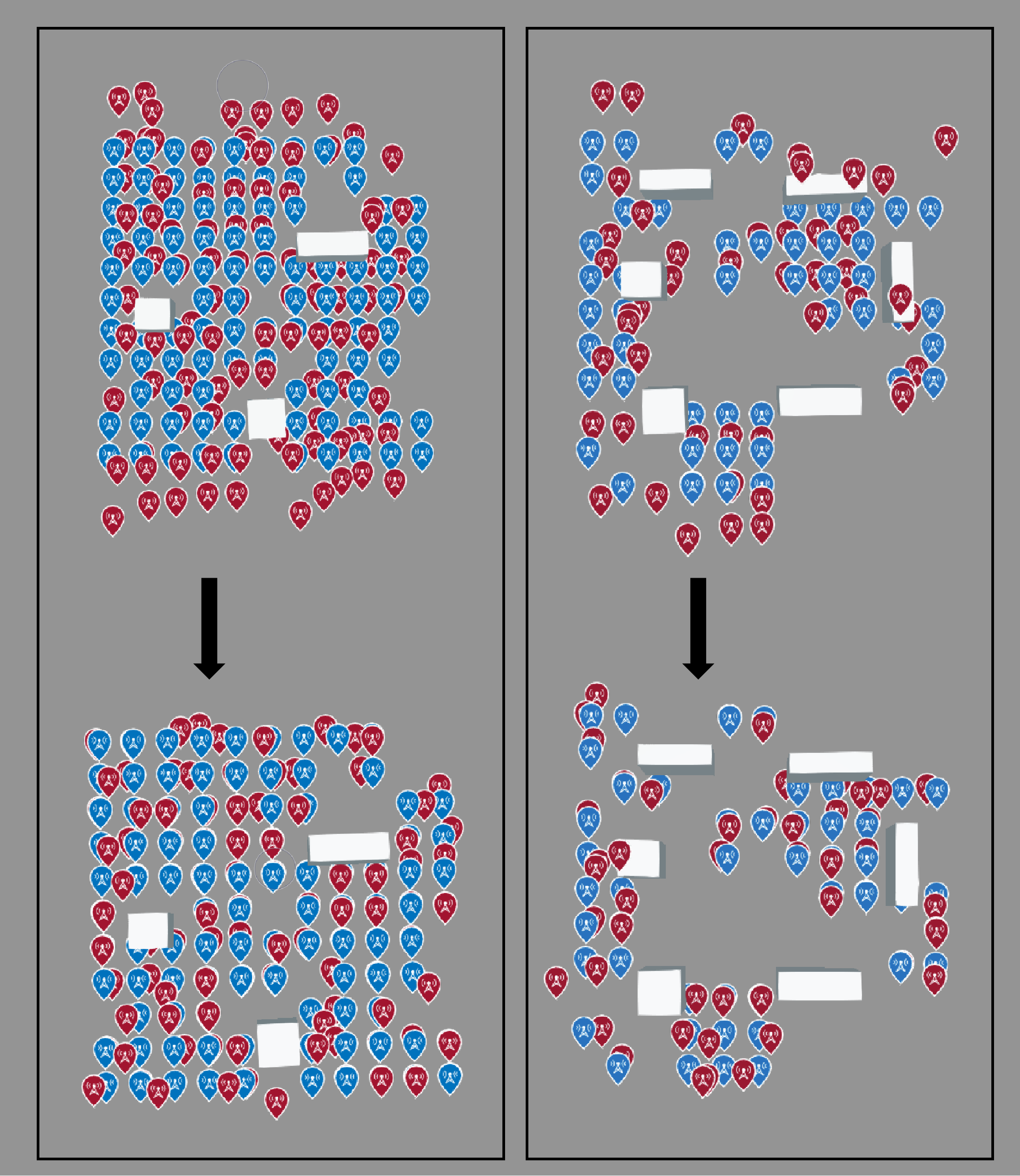}}
\end{minipage} 
\caption{Error compensation performance of the CNN localization error prediction.}
\label{fig7}
\end{figure}

\setlength{\parindent}{1em}As exemplified in Fig.\ref{fig7}, the localization DA-error predicted by CNN is used to correct the results obtained from TDOA. The blue points represent the ground truth positions, while the red points depict the location distributions of TDOA or those after localization DA-error compensation. When more buildings are distributed, including additional NLOS propagation mode, fewer UE locations can be directly determined, leading to larger localization errors with TDOA alone. When only TDOA is applied, as shown in the top half of Fig.\ref{fig7}, the localization results significantly deviate from the ground truth. The red dots representing TDOA results are far from the ground truth (blue dots), and their distribution is more scattered, especially near the map's edges or regions blocked by buildings. After the localization DA-error correction, corresponding to the bottom half of Fig.\ref{fig7}, it is evident that the positions (red dots) after correction overlap more with or are closer to the ground truth compared to those before error correction. 

\begin{figure}[t]
	\includegraphics[width=3.5in]{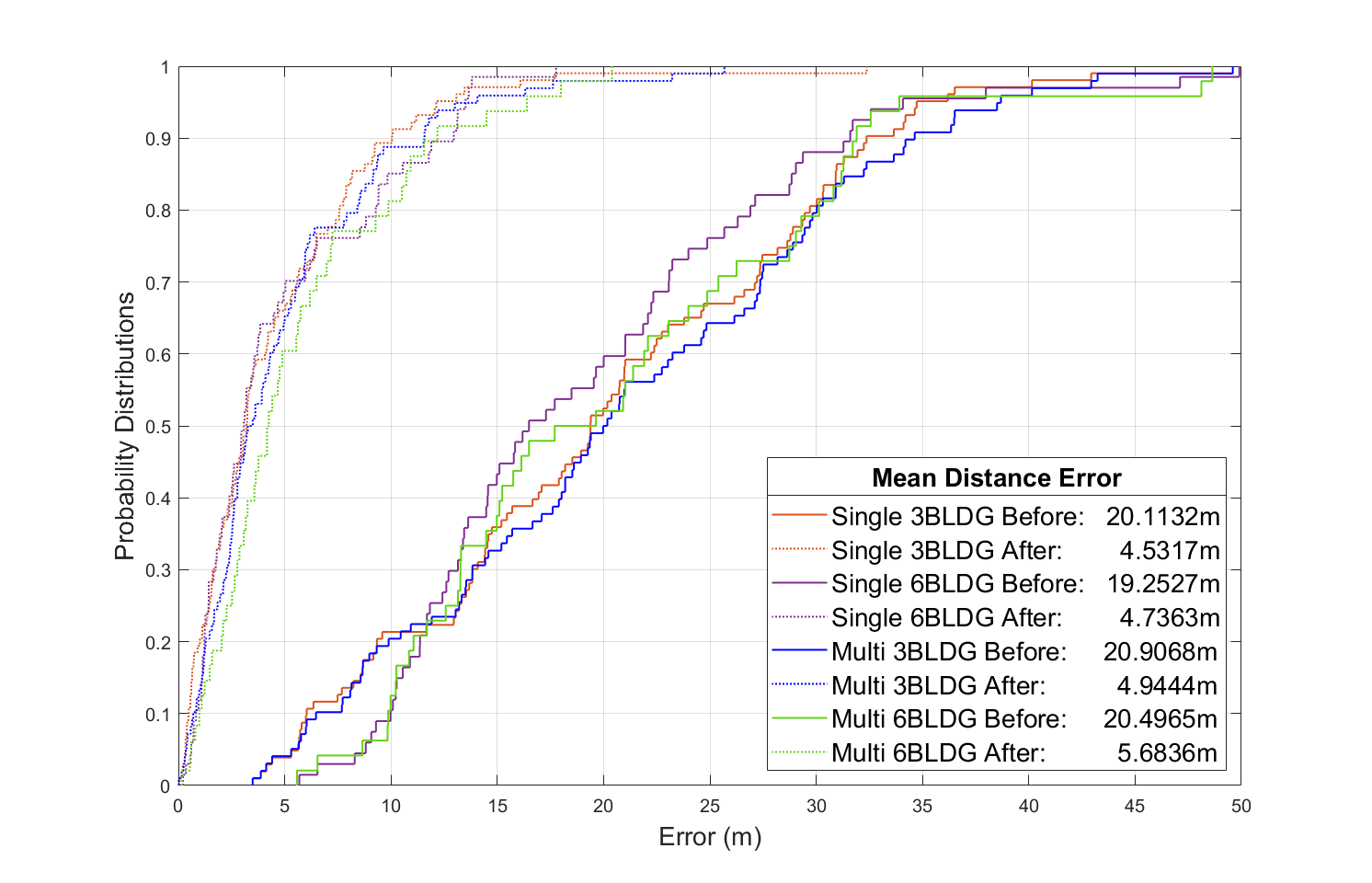}
\caption{Probability distributions performance of the error compensation.}
\label{fig8}
\end{figure}

\setlength{\parindent}{1em}Fig.\ref{fig8} shows the error probability distributions and statistical results of localization performance, i.e., the averaged Euclidean distances between the ground truth and the localization results. It indicates that: (1) In the NLOS environment, the localization performance of single path is better than that of multi-path, which is in accordance with the multi-path effect in NLOS environment. (2) The more buildings are distributed in the map, the stronger interference to the localization results will exit after DA-error compensation. (3) In general, after correcting the localization error, the median error of TDOA was reduced from about 20m to around 5m, corresponding to about 75\% gain in accuracy by the proposed CNN-based localization enhancement scheme.

\setlength{\parindent}{1em}The simulation results above demonstrate that CNN exhibits high reliability in predicting DA-error information, and the accuracy significantly improves after error compensation. This effectively validates the positioning enhancement scheme using CNN. 

\section{Conclusion}
In order to improve the localization accuracy of TDOA, this paper introduces CNN to extract multi-path features from maps predicting localization DA-errors, which are then used to compensate the TDOA results affected by NLOS propagation. Firstly, the motivation for proposing CNN to predict TDOA localization DA-error is analyzed. Secondly, a new localization correction scheme is designed, compensating for the predicted error. 4 tasks of localization DA-error prediction are carried out to train CNN under different scenarios. Finally, a comparative analysis is provided for TDOA localization accuracy before and after DA-error compensation. According to the results, for the propagation environment with both LOS and NLOS propagation modes, the performance of TDOA is significantly affected by building occlusion and multi-path interference. The proposed DA-error prediction method substantially improves localization accuracy. As for the future work, the proposed scheme might be extended to the case where a UE is moving along certain trajectories, exploiting the potential spatial correlation among errors extracted by CNN. Besides, data from other sensors, such as an inertial measurement unit, can be combined to enhance localization accuracy. The application of other AI models, such as ResNet or Transformer, might also be considered in this study.


\begin{thebibliography}{1}
\bibitem{b1}
Compagnoni, M., Canclini, A., Bestagini, P. \textit{et al.} ``Source localization and denoising: a perspective from the TDOA space,'' \textit{Multidim Syst Sign Process 28}, 1283–1308 (2017).

\bibitem{b2}
G. Fokin, ``TDOA Measurement Processing for Positioning in Non-Line-of-Sight Conditions,'' \textit{2018 IEEE International Black Sea Conference on Communications and Networking (BlackSeaCom)}, Batumi, Georgia, 2018, pp. 1-5.

\bibitem{b3}
X. Zhou, C. Xu, J. He and J. Wan, ``A Cross-region Wireless-synchronization - based TDOA Method for Indoor Positioning Applications,'' \textit{2019 28th Wireless and Optical Communications Conference (WOCC)}, Beijing, China, 2019, pp. 1-4.

\bibitem{b4}
L. Huang \textit{et al.}, ``Robust TDOA-Based Indoor Localization Using Improved Clock-Sync-Scheme and Multilevel Constrained ARPF,'' in \textit{IEEE Sensors Journal}, vol. 23, no. 10, pp. 10633-10643, 15 May15, 2023.

\bibitem{b5}
M. Han and G. Zeng, ``Research on indoor radio frequency positioning algorithm based on TDOA,'' \textit{2022 International Seminar on Computer Science and Engineering Technology (SCSET)}, Indianapolis, IN, USA, 2022, pp. 89-92.

\bibitem{b6}
Y. Xue, W. Su, H. Wang, D. Yang and Y. Jiang, ``DeepTAL: Deep Learning for TDOA-Based Asynchronous Localization Security With Measurement Error and Missing Data,'' in \textit{IEEE Access}, vol. 7, pp. 122492-122502, 2019.

\bibitem{b7}
J. Cho, D. Hwang and K. -H. Kim, ``Improving TDoA Based Positioning Accuracy Using Machine Learning in a LoRaWan Environment,'' \textit{2019 International Conference on Information Networking (ICOIN)}, Kuala Lumpur, Malaysia, 2019, pp. 469-472.

\bibitem{b8}
A. Niitsoo, T. Edelhäußer and C. Mutschler, ``Convolutional Neural Networks for Position Estimation in TDoA-Based Locating Systems,'' \textit{2018 International Conference on Indoor Positioning and Indoor Navigation (IPIN)}, Nantes, France, 2018, pp. 1-8.

\bibitem{b9}
J. Son, I. Keum, Y. Ahn and B. Shim, ``D-TLoc: Deep Learning-aided Hybrid TDoA/AoA-based Localization,'' \textit{2022 IEEE VTS Asia Pacific Wireless Communications Symposium (APWCS)}, Seoul, Korea, Republic of, 2022, pp. 47-50.

\bibitem{b10}
G. Kia, L. Ruotsalainen and J. Talvitie, ``A CNN Approach for 5G mm Wave Positioning Using Beamformed CSI Measurements,'' \textit{2022 International Conference on Localization and GNSS (ICL-GNSS)}, Tampere, Finland, 2022, pp. 01-07.

\bibitem{b11}
H. Cheng, S. Ma and H. Lee, ``CNN-Based mmWave Path Loss Modeling for Fixed Wireless Access in Suburban Scenarios,'' in \textit{IEEE Antennas and Wireless Propagation Letters}, vol. 19, no. 10, pp. 1694-1698, Oct. 2020


\bibitem{b12}
H. Cheng, S. Ma, H. Lee and M. Cho, ``Millimeter Wave Path Loss Modeling for 5G Communications Using Deep Learning With Dilated Convolution and Attention,'' in \textit{IEEE Access}, vol. 9, pp. 62867-62879, 2021

\bibitem{b13}
L. Zang, C. Shen, K. Zhang, L. Xu and Y. Chen, ``Research on Hybrid Algorithm Based on TDOA,'' \textit{2020 IEEE 20th International Conference on Communication Technology (ICCT)}, Nanning, China, 2020, pp. 539-542.

\bibitem{b14}
L. Chang-jiang, W. Chao-feng and L. Dan, ``Influence of Correction for Atmospheric Refraction in passive TDOA location on system TDOA error,'' \textit{2019 International Conference on Control, Automation and Information Sciences (ICCAIS)}, Chengdu, China, 2019, pp. 1-4.

\bibitem{b15}
Y. Chen, T. Xiang and X. Zhang, ``An Efficient Wireless Propagation Loss Prediction Model Based on 3-D Terrain Features Extracted by Deep Learning,'' in \textit{IEEE Antennas and Wireless Propagation Letters}, vol. 22, no. 5, pp. 1055-1058, May 2023.

\bibitem{b16}
S. Zelong, Q. Cheng, J. Weiwei, X. Xianwen, L. Chao and O. Shangrong, ``Amplitude and phase error estimation and compensation technology for spaceborne SAR,'' \textit{2022 3rd China International SAR Symposium (CISS)}, Shanghai, China, 2022, pp. 1-4.

\bibitem{b17}
S. Sellami and A. Klimchik, ``A deep learning based robot positioning error compensation,'' \textit{2021 International Conference "Nonlinearity, Information and Robotics" (NIR)}, Innopolis, Russian Federation, 2021, pp. 1-5.

\bibitem{b18}
Yiwen Chen, Tianqi Xiang, Xi Chen, Xin Zhang, September 27, 2023, ``Map and Localization error prediction dataset'', IEEE Dataport, doi: https://dx.doi.org/10.21227/1b3x-ca08.

\bibitem{b19}
Simonyan K, Zisserman A. Very deep convolutional networks for large-scale image recognition[J]. arXiv preprint arXiv:1409.1556, 2014.

\bibitem{b20}
``Study on NR positioning support'', \textit{3rd Generation Partnership Project (3GPP) Tech. Rep. 38.855}, 2019, [online] Available: https://www.3gpp.org/ftp/Specs/archive/38\_series/38.855/ 38855-g00.zip.

\end{thebibliography}
\end{document}